\documentclass[aps,prl,twocolumn, superscriptaddress,showpacs,floatfix]{revtex4-1}
\usepackage{float}
\usepackage{graphicx}
\usepackage{amsmath}
\usepackage{amssymb}
\usepackage{url}
\usepackage{color}
\usepackage{natbib}

\usepackage{bm}

\graphicspath{{.}{../}{./eps/}}

\renewcommand{\v}[1]{\ensuremath{\mathbf{#1}}} 

\renewcommand{\d}[2]{\frac{d #1}{d #2}} 

\newcommand{\ket}[1]{\left| #1 \right>} 
\newcommand{\bra}[1]{\left< #1 \right|} 
\let\baraccent=\= 
\renewcommand{\=}[1]{\stackrel{#1}{=}} 

\newcommand{\eq}[1]{(\ref{#1})}

\newcommand{\medio}[1]{\left\langle #1 \right\rangle}

\def\XXint#1#2#3{{\setbox0=\hbox{$#1{#2#3}{\int}$}
     \vcenter{\hbox{$#2#3$}}\kern-.5\wd0}}

\renewcommand{\exp}[1]{\text{e}^{#1}}

\begin{document}

\title{ Real-space calculation of the conductivity tensor for disordered topological matter}

\author{Jose H. Garc\'ia}
\affiliation{Instituto de F\'isica, Universidade Federal do Rio de Janeiro, Caixa Postal 68528, 21941-972 Rio de Janeiro RJ, Brazil}

\author{Lucian  Covaci}
\affiliation{Department Fysica, Universiteit Antwerpen, Groenenborgerlaan 171, B-2020 Antwerpen, Belgium}

\author{Tatiana G. Rappoport}
\affiliation{Instituto de F\'isica, Universidade Federal do Rio de Janeiro, Caixa Postal 68528, 21941-972 Rio de Janeiro RJ, Brazil}

\date{\today}

\begin{abstract}
We describe an efficient numerical approach to calculate the longitudinal and transverse Kubo conductivities of large systems using Bastin's formulation \cite{Bastin}. We expand the Green's functions in terms of Chebyshev polynomials and compute the conductivity tensor for any temperature and chemical potential in a single step. To illustrate the power and generality of the approach, we calculate the conductivity tensor for the quantum Hall effect in disordered graphene and analyze the effect of the disorder in a Chern insulator in Haldane's model on a honeycomb lattice.
\end{abstract}


\maketitle


One of the most important experimental probes in condensed matter physics is the electrical response  to an external electrical field. In addition to the longitudinal conductivity, in specific circumstances, a system can present a transverse conductivity under an electrical perturbation. The Hall effect~\cite{HE} and the anomalous Hall effect in magnetic materials~\cite{AHE} are two examples of this type of response. Paramagnetic materials with spin-orbit interaction can also present transverse spin currents~\cite{SHE}.
There are also the quantized versions of the three phenomena: while the quantum Hall effect (QHE) was observed more than 30 years ago~\cite{QHE}, the quantum spin Hall effect (QSHE) and the quantum anomalous Hall effect (QAHE)  could only be observed~\cite{QAHE,QSHE} with the recent discovery of topological insulators, a new class of quantum matter~\cite{topo}.

In the linear response regime, the conductivity tensor can be calculated using the Kubo formalism
~\cite{Kubo}. The Hall conductivity can be easily obtained in momentum space in terms of the Berry
curvature associated with  the bands \cite{thouless}. The downside of working in momentum space,
however, is that the robustness of a topological state in the presence of disorder can only be
calculated  perturbatively ~\cite{ando}. Real-space implementations of the Kubo formalism for the 
Hall conductivity, on the other hand, allow the incorporation of different types of disorder in varying degrees, 
while providing flexibility to treat different geometries.  Real-space techniques, however, normally
require a large computational effort. This has generally restricted their use to either small
systems at any temperature \cite{exactd, prodan}, or large systems at zero temperature
\cite{roche}. 

In this Letter, we propose a new efficient numerical approach to calculate the conductivity tensor
in solids. We use a real space implementation of the Kubo formalism where both diagonal and
off-diagonal conductivities are treated in the same footing. We adopt a formulation of the
Kubo theory that is known as Bastin formula~\cite{Bastin} and expand the Green's functions involved
in terms of Chebyshev polynomials using the kernel polynomial method~\cite{KPM}.
There are  few numerical methods that use Chebyshev expansions to calculate
the longitudinal DC conductivity ~\cite{roche_old, aires,harju,katsnelson} and 
transverse conductivity \cite{roche,verga} at zero temperature. An advantage of our approach is the possibility of obtaining 
both conductivities for large systems in a single calculation step, independently of the temperature, chemical potential and 
for any amount of disorder.

We apply this method to two different systems displaying topological states in a honeycomb
lattice. The first one has been extensively explored in the literature~\cite{RMPgraphene,
RMPdSarma,roche,mirlin}, and consists of disordered graphene under constant perpendicular
magnetic field. Our calculation of the longitudinal and Hall conductivities serve to
illustrate the key aspects of our approach.  We then apply the method to  a Chern insulator (CI) 
in Haldane's model on a honeycomblattice~\cite{Haldane}. This model produces an
 insulating state with broken time-reversal symmetry in the absence of a macroscopic magnetic field.
Instead of behaving as a
normal insulator, it  exhibits a quantized Hall conductivity $\sigma_{xy}=e^2/h$ in the gapped
state.  If the inversion symmetry is broken, the system can undergo a topological phase transition
to a normal insulator. We investigate the transport properties of Chern insulators and analyze how
they are affected by the interplay between disorder and inversion symmetry breaking.

The conductivity tensor can be calculated using the Kubo formula from linear response theory. In the limit $\omega\rightarrow 0$, the elements of the static conductivity tensor for non-interacting electrons are given by the Kubo-Bastin formula for the conductivity~\cite{Bastin}

\begin{align}
&\tilde{\sigma}_{{\alpha\beta}}(\mu, T)=\frac{ie^2\hbar}{\Omega}
\int_{-\infty}^{\infty}d\varepsilon f(\varepsilon) \label{DCconductivity}\\
&\times \text{Tr}\left\langle v_\alpha\delta(\varepsilon-H)v_\beta \d{G^{+}(\varepsilon)}{\varepsilon}-v_\alpha  \d{G^{-}(\varepsilon)}{\varepsilon}v_\beta \delta(\varepsilon-H)\right \rangle ,   \nonumber
\end{align}
where {$\Omega$ is the volume}, $v_\alpha$  is the  $\alpha$ component of the velocity
operator, $G^{\pm}(\varepsilon,H)=\frac{1}{\varepsilon-H\pm i0}$ are the advanced $(+)$ and retarded
$(-)$ Green's functions,
 and $f(\varepsilon)$ is the Fermi-Dirac distribution for a given temperature $T$ and chemical potential $\mu$.
The expression above was first obtained by Bastin and collaborators in 1971~\cite{Bastin} and later
generalized for any independent electron approximation~\cite{Bruno}. However, it has not been used often
in numerical calculations because of the complications of dealing with an integration in energy.
Instead, it is possible to perform analytical integrations by parts~\cite{Bruno} to obtain a more
treatable expression for the static conductivity at zero temperature, which became known as the
Kubo-Streda formula~\cite{Streda_formula}. For the diagonal elements of the conductivity tensor ($\alpha=\beta$),
the integration leads to the
Kubo-Greenwood formula~\cite{Greenwood}.  

{Here we propose a new approach to compute, for any finite temperature, both diagonal and
off-diagonal conductivities using the Kubo-Bastin formula}. Our method consists in expanding the
Green's functions in the integrand of eq. (\ref{DCconductivity}) in terms of Chebyshev polynomials
using the kernel polynomial method~\cite{KPM,KPM1}, a highly efficient and scalable way to calculate
the Green's functions in electronic systems \cite{KPM,KPMgreen1Lucian,KPMgreen2,KPMgreen3}. For that purpose,
we first need to rescale the Hamiltonian so that the upper $E^+$ and  lower $E^-$ bounds of the
spectrum  are mapped into 1 and -1 respectively. {To estimate the bounds, we apply the power method~\cite{powerbook}, which is normally used to locate dominant eigenvalues in linear algebra}. The rescaled Hamiltonian and energy are represented
by $\tilde{H}$ and $\tilde{\varepsilon}$~\cite{SM} and we can expand the rescaled delta and Green's
functions by considering their spectral representations  and expanding their eigenvalues in terms of
the Chebyshev polynomials:
\begin{align}
&\delta(\tilde{\varepsilon}-\tilde{H})=\frac{2}{\pi \sqrt{1-\tilde{\varepsilon}^2}}\sum_{m=0}^M g_m\frac{T_m(\tilde{\varepsilon})}{\delta_{m,0}+1}T_m(\tilde{H}), \\
& G^\pm(\tilde{\varepsilon},\tilde{H})=\mp \frac{2i}{\sqrt{1-\tilde{\varepsilon}^2}}\sum_{m=0}^M g_m\frac{\exp{\pm i m \text{arccos}(\tilde{\varepsilon})}}{\delta_{m,0}+1} T_m(\tilde{H}).
\end{align}
where {$T_m(x)=\cos[m\arccos(x)]$} is the Chebyshev polynomial of the first kind and order
$m$, which is defined according to the recurrence relation
{$T_{m}(x)=2xT_{m-1}(x)-T_{m-2}(x)$}. The expansion has a finite number of terms ($M$) and  the
truncation gives rise to Gibbs oscillations that can be smoothed with the use of a kernel,
given by $g_m$~\cite{KPM,KPM1}.

Replacing the expansions above in \eq{DCconductivity} with $\Delta E=E^{+}-E^-$,  we obtain
\begin{align}
\sigma_{{\alpha\beta}}(\mu, T)&= \frac{4e^2\hbar}{\pi \Omega}\frac{4}{\Delta E^2}
\int_{-1}^{1}d\tilde{\varepsilon}
\frac{f(\tilde{\varepsilon})}{(1-\tilde{\varepsilon}^2)^2}\sum_{m,n}\Gamma_{{nm}}
(\tilde { \varepsilon } )\mu^{\alpha\beta}_{{nm}}\label{condfinal}
\end{align}
where $\mu^{\alpha\beta}_{{mn}}\equiv \frac{g_m
g_n}{(1+\delta_{{n0}})(1+\delta_{{m0}})}\text{Tr}\left[v_\alpha T_m(\tilde{H})v_\beta
T_n(\tilde{H})\right]$ does not depend on the energy. Since $\mu_{{mn}}$ involves products of
polynomial expansions of the Hamiltonian, its calculation is responsible for most of the method's
computational cost.

On the other hand, {$\Gamma_{mn}(\tilde{\varepsilon})$} is a scalar that is energy dependent
but independent of the Hamiltonian
\begin{equation}
\begin{split}
{\Gamma_{mn}(\tilde{\varepsilon})\equiv[(\tilde{\varepsilon}-i
n\sqrt{1-\tilde{\varepsilon}^2})\exp{in\text{arccos}(\tilde{\varepsilon})}T_m(\tilde{\varepsilon})}
\\
+(\tilde{\varepsilon}+im\sqrt{1-\tilde{\varepsilon}^2})\exp{-im\text{arccos}(\tilde{\varepsilon})}T_n(\tilde{\varepsilon})].
\end{split}
\end{equation}
As shown in \eq{condfinal}, once the coefficients $\mu_{{mn}}$ are determined, we can obtain
the conductivities for all temperatures and chemical potentials without repeating the most
time-consuming part of the calculation~\cite{kpm_accond}. Moreover, the recursive relations between
Chebyshev polynomials lead to a recursive multiplication of sparse Hamiltonian matrices that can be
performed in a very efficient way in GPUs~\cite{KPMgreen1Lucian,harju}.  Instead of the full
calculation of traces, we use self-averaging properties, normally used in Monte-Carlo calculations,
to replace the trace in the calculation of  $\mu_{{mn}}$ by the average of a small number $R
\ll N$ of random phase vectors $|r\rangle$ and  further improve the efficiency of the
calculation~\cite{RandomVector1,RandomVector2}.  The conductivities are averaged over several
disorder {realizations}, $S$, with $R=5$ for each of them. Because of the self-averaging properties of large systems,
 the product $SR$ is the main defining factor of the accuracy of the trace operation.

The first problem we apply our method to is the physics of the QHE in disordered graphene.  We start from the electronic Hamiltonian of graphene in the presence of a random scalar potential and a perpendicular magnetic field $\mathcal{H}=-t\!\sum_{\medio{i,j}} e^{i\phi_{ij}}c^\dagger_i c_j +\sum_{i}\varepsilon_i c^\dagger_i c_i
$ where $c_i$  is  the annihilation operator of electrons on  site $i$ where $t\approx$ 2.8 eV is
the hopping energy between nearest neighbors (NN) sites in a honeycomb lattice. The  perpendicular
magnetic field is included by Peierls' substitution $\phi_{ij}=2\pi/\Phi_0\int_j^i \vec{A}\cdot d
\vec{l}$. Using the Landau gauge $\vec{A}=(-By,0,0)$,  the phase will be $\phi_{ij}=0$ along the $y$
direction and  $\phi_{ij}=\pm \pi (y/a) \Phi/\Phi_0$ along the $\mp x$ direction,  where $\Phi$ is
the magnetic flux per unit cell, $\Phi_0$ being the quantum of magnetic flux. The second term in
$\mathcal{H}$ represents the on-site Anderson disorder where $\varepsilon_i$ is randomly chosen from
a uniform probability distribution $p(\varepsilon_i)=\frac{1}{\gamma}\theta\left(\frac{\gamma}{2}
-|\varepsilon_i|\right)$, where $\gamma$ accounts for the amount of disorder introduced in the
system.
 \begin{figure}[b]
     \centering\includegraphics[width=0.9\columnwidth,clip]{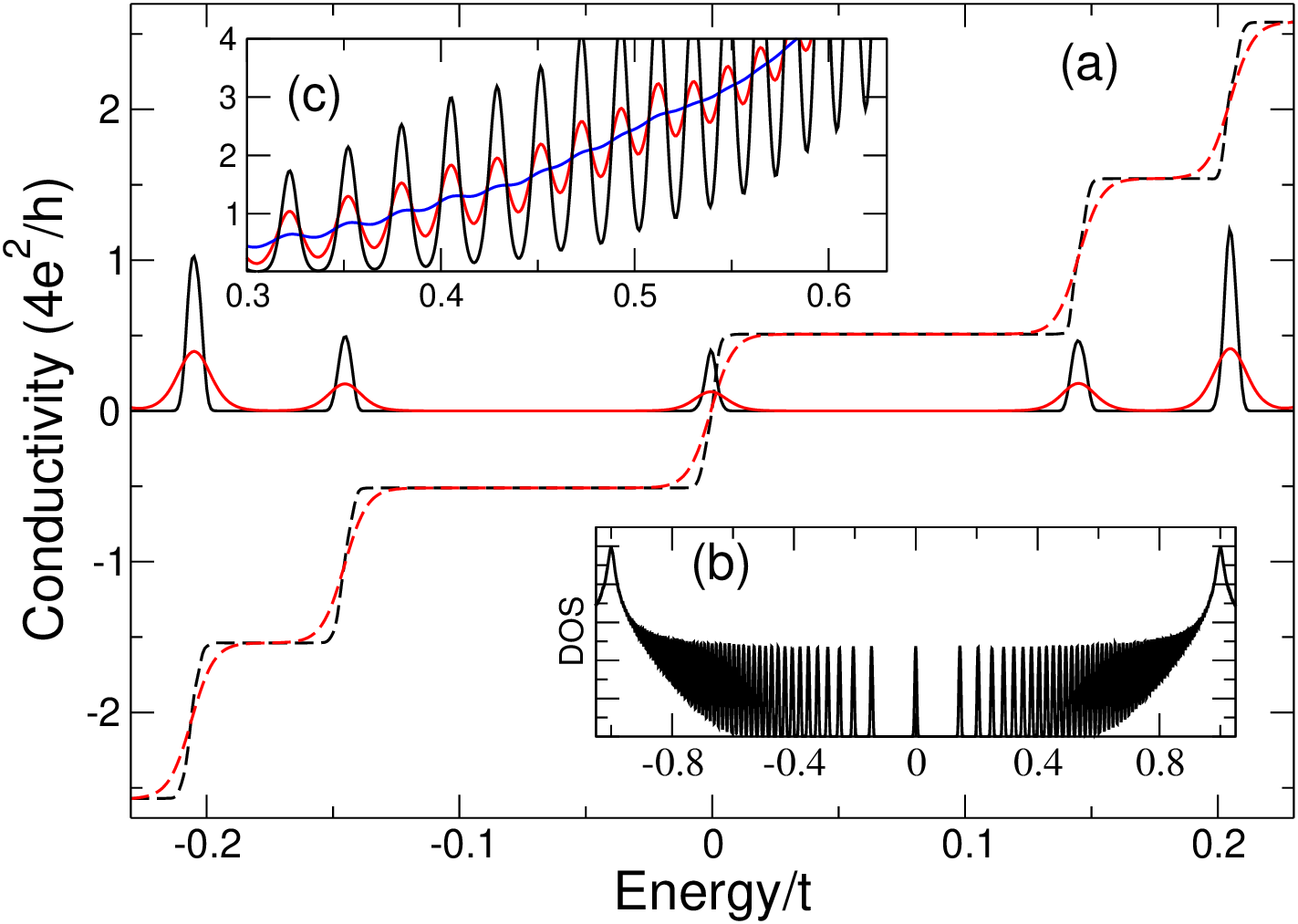}
     \caption{ (a)  $\sigma_{xx}$ (solid line) and  $\sigma_{xy}$ (dashed line) for $k_B T=0$ (black) and $k_BT/t=0.004$ (red).  (b) Electronic density of states  (c) $\sigma_{xx}$ away from the Dirac point where  Shubnikov-de Haas oscillations can be observed for $k_BT/t=0.002$  (black) , $k_BT/t=0.004$  (red), $k_BT/t=0.008$  (blue) . The parameters in panels (a)-(c) are  $\phi/\phi_0\approx 1\times 10^{-3}$, $SR=200$, $M=6144$ and $N=2\times 128\times 1024$ sites where we use a rectangular geometry to minimize the magnetic flux per unit cell in a system with periodic boundary conditions. }
     \label{fig:qhe}
\end{figure}
Let us begin with  a graphene layer with $N\approx 2.6 \times 10^5$ sites with periodic boundary conditions and weak disorder given by $\gamma=0.1 t$ and $SR=200$.  
In the presence of a perpendicular magnetic field such that $\Phi/\Phi_0\approx 1\times 10^{-3}$, the electronic density of states (DOS) presents several Landau levels close to the Dirac point. Away from $E=0$, the magnetic length is larger than the system size; the band structure still presents a large number of peaks, with a non-zero density of states between the peaks, which results in a metal behavior, as seen in  Fig. \ref{fig:qhe} (b). We compute the longitudinal and off-diagonal conductivities as a function of the chemical potential $\mu$ and close to $\mu=0$ the results are consistent with the QHE in pure graphene. Figure \ref{fig:qhe} (a) shows the peaks in  $\sigma_{xx}$ that are located exactly at the peaks of the density of state. For $\sigma_{xx}=0$  we see well-resolved  plateaus of the Hall conductivity following $\sigma_{xy}=4e^2/h(n+1/2)$, indicating that the method captures the topological nature of the insulating phase. The effect of the temperature is the predictable broadening of the longitudinal conductivity peaks together with the smearing of the quantum Hall plateau. Fig. \ref{fig:qhe} (c) reports  Shubnikov-de Haas oscillations in the longitudinal conductivity away from the Dirac point. Similarly to what is observed experimentally~\cite{novoselov05}, they are sensitive to changes in $T$.
\begin{figure}[!h]
    \centering\includegraphics[width=0.9\linewidth,clip]{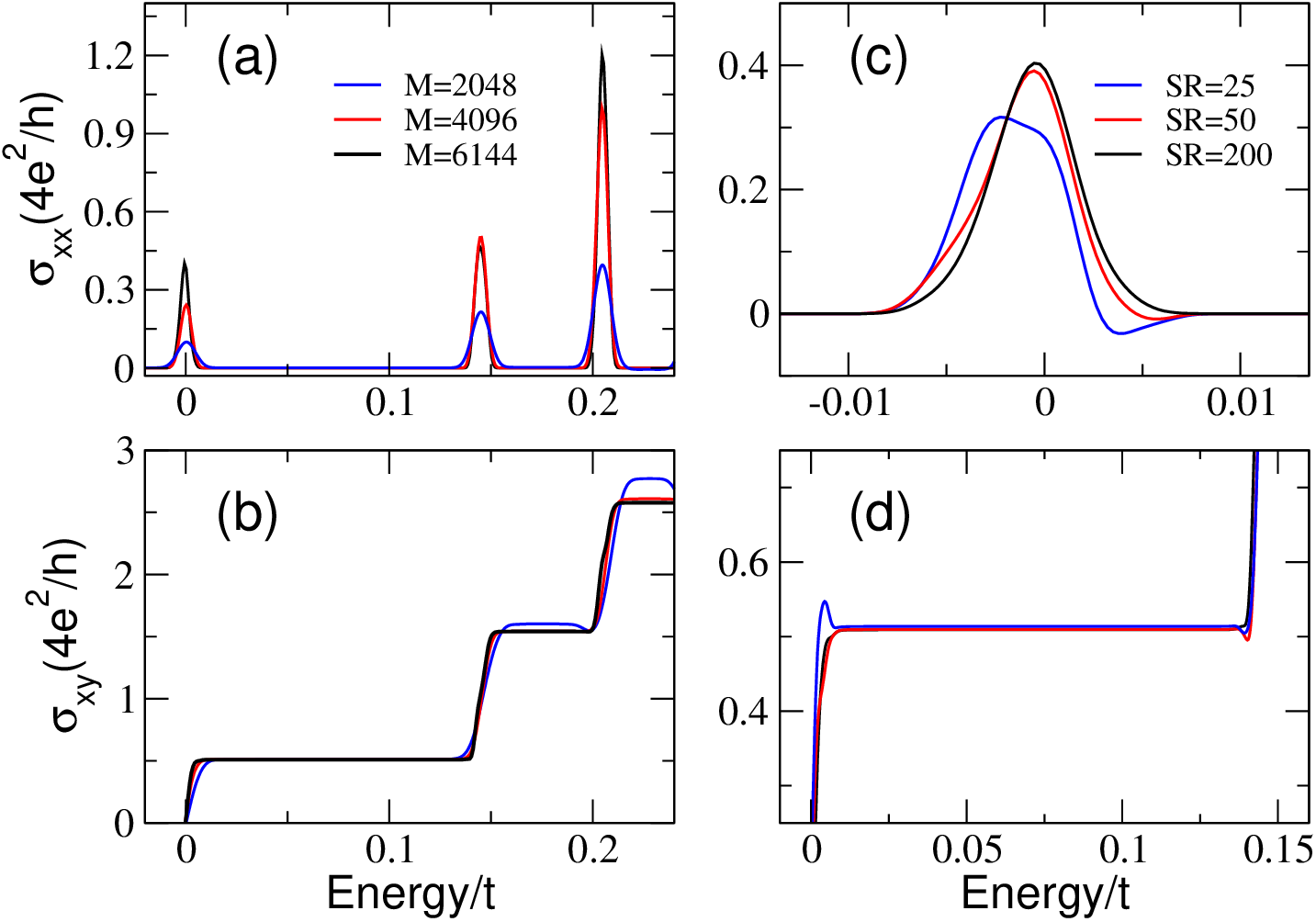}
     \caption{ Analysis of the dependency on the number of random vectors and disorder realizations $SR$ and moments $M$: (a)  $\sigma_{xx} $  and (b) $\sigma_{xy}$ for $SR=200$ and $M=2048$ (blue), 4096 (red) and 6144 (black).  (c)  $\sigma_{xx}$   and  (d) $\sigma_{xy}$ for $M=6144$ and $SR=25$ (blue), 50 (red) and 200 (black).  In panels (a)-(d) ,  $\phi/\phi_0=5\times 10^{-3}$,  and $N=2\times 128\times 1024$ sites  }
     \label{fig:conv}
\end{figure}
To get results as accurate as those in Fig. \ref{fig:qhe}, one needs to look at the convergence of the expansion as a function of the few parameters that were introduced in our approach, such as the polynomial order $M$ and  the product $SR$. To illustrate this, in Fig. \ref{fig:conv} we show the dependence of $\sigma_{xx}$ and $\sigma_{xy}$ on $M$ and  $SR$. For disordered systems, the interference due to quasi-particle scattering off the impurities~\cite{KPMgreen1Lucian} results in an oscillatory behavior of the Chebyshev moments. Because of this, an accurate solution requires a large number of moments. {The energy resolution of the KPM depends on $M$ and its value is important for the convergence of the sharp peaks of $\sigma_{xx}$. This is illustrated in Fig.~\ref{fig:conv} (a) where the conductivity peak at $E=0$ is consistent with recent numerical calculations~\cite{roche,mirlin} and its convergence is only achieved for $M>6000$.

The energies of the Laudau levels close to the Dirac point scale with $\sqrt{n}$, reducing the gap between high Landau levels. Simultaneously, the density of states increases with $E$. Consequently,  we need more moments in the expansion to resolve small gaps and localize carriers in regions of the spectra with more available states.  As can be seen in  Fig.~\ref{fig:conv} (b), this results in a non-homogeneous convergence of the expansion:  the plateaus located close to $E=0$ converge for lower values of $M$ while the higher Landau levels need more moments to converge.  To ensure accurate results, we can track the global convergence of the conductivity as a function of $M$  in a desirable energy window~\cite{SM}.} 

We also need a large $SR$ to achieve the self-averaging condition~\cite{KPM}. In particular, $\sigma_{xx}$ and the transition between quantum Hall plateaus are sensitive to $SR$ as illustrated in Fig. \ref{fig:conv} (c) and (d) and convergence is obtained for $SR>125$. From Fig. \ref{fig:conv}, we can see that intermediate values of $M$ and $SR$ are enough for a qualitative analysis of $\sigma_{\alpha\beta}$. {For higher accuracy one needs larger values of $M$, which for good convergence would also require $SR$ to be increased.}

Non-trivial topologies in the band structure can also occur in the absence of an external magnetic field. In Chern insulators, time-reversal symmetry is explicitly broken without the need of an external magnetic field. In this sense, these systems can be seen as the quantized version of the AHE that has been recently observed experimentally~\cite{chern_science}.  A simple model proposed by Haldane \cite{Haldane} in a honeycomb lattice provides all the key ingredients of Chern insulators. The Hamiltonian is
\begin{align}
\mathcal{H}=-t\!\sum_{\medio{i,j}} c^\dagger_i c_j+ t_2\!\sum_{ \langle \medio{i,j} \rangle} e^{i\phi_{ij}}c^\dagger_i c_j\pm\frac{\Delta_{AB}}{2}\sum_{i\in A/ B }c^\dagger_i c_i , \label{Hhaldane}
\end{align}
where $t$ and $t_2$ are nearest and next-nearest-neighbor hopping amplitudes. $\phi_{ij}$  is equivalent to a Peierls phase with zero total flux per unit cell. The last term is an energy offset between sublattices $A$ and $B$ that breaks the inversion symmetry of the Hamiltonian, opening a gap $\Delta_{AB}$ in the band structure. For $\phi=\pi/2$ and $\Delta_{AB}=0$, the system also presents a gap of $ \Delta_T=6\sqrt{3} t_2$,  and if $\mu$ lies inside the gap, the system is a Chern insulator with $\sigma_{xy}=e^2/h$. If $\Delta_{AB} $ is continuously increased, it undergoes a quantum phase transition from a  Chern insulator to a normal insulator for $\Delta_{AB}>\Delta_T$~\cite{Haldane}.
\begin{figure}[h]
     \centering\includegraphics[width=0.95\linewidth,clip]{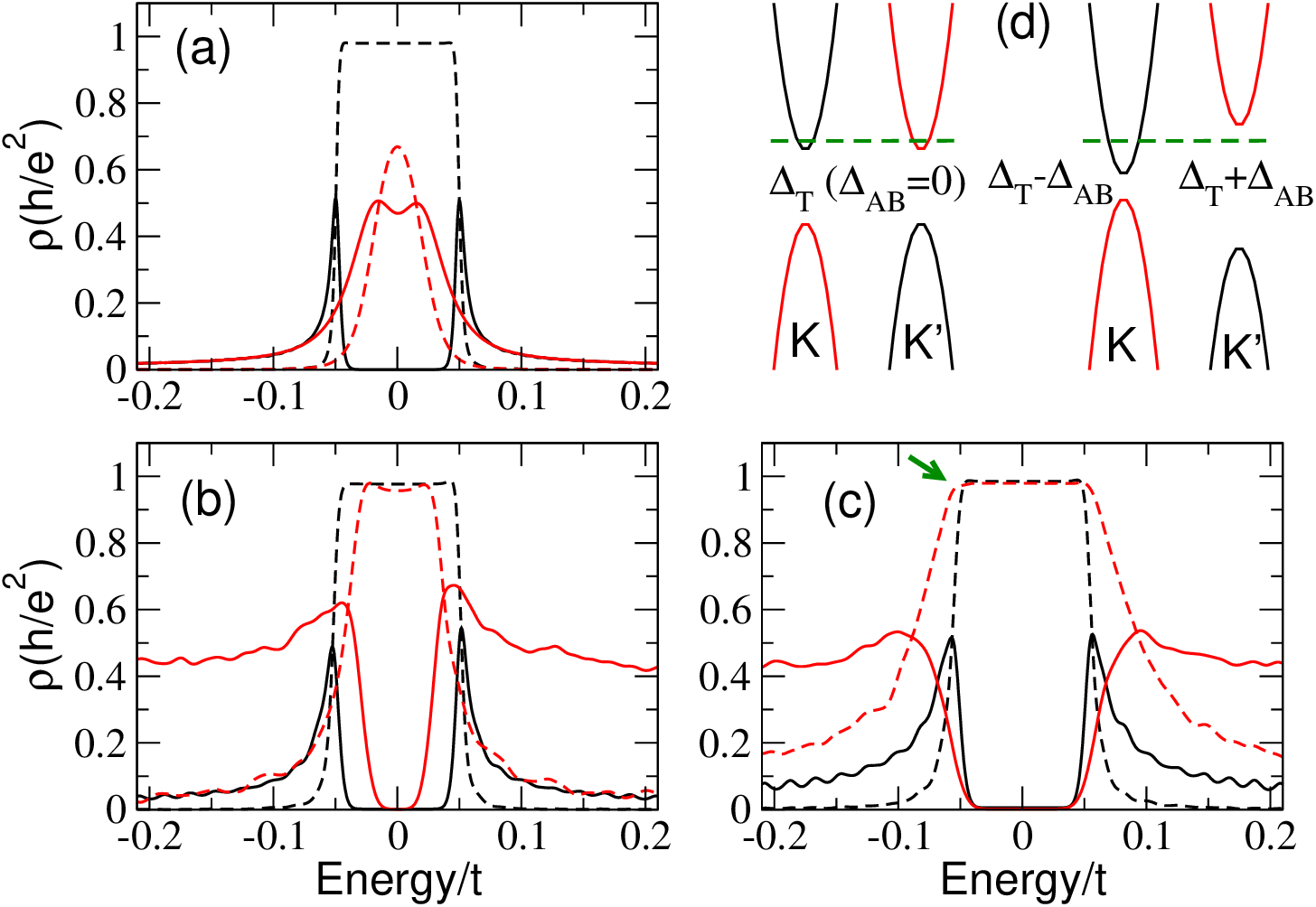}
     \caption{  $\rho_{xx}$ (solid line) and $\rho_{xy}$ ({dashed} line) for HM with  (a)
$\Delta_T=0.1t$,  $\Delta_{AB}=0$ and $\gamma=0.2 t$ for $k_B T=0$ (black) and $k_BT=0.16t$ (red);
(b)  $\Delta_T=0.1 t$,  $\Delta_{AB}=0$ and $k_B T=0$ for $\gamma=0.4 t$(black) and $\gamma=1.8 t$
(red); (c)  $\Delta_T=0.5 t$,  $\Delta_{AB}=0.4t$ and $k_B T=0$ for $\gamma=0.2 t$ (black) and
$\gamma=1.8 t$ (red). The green arrow indicates the increase of the topological region with
disorder.  The system sizes in panels  $D=2\times L\times L$ with $L=512$  (a), $L=256$ (b-c) and  $SR>200$  . Panel
(d) illustrates the different gap sizes at $K$ and $K^\prime$ for $\Delta_{AB}\neq 0$.}
     \label{fig:haldane}
\end{figure}
We proceed to investigate the QAHE  for $\Delta_{AB}=0$ in the presence of Anderson disorder with bounds $\pm\gamma$.  As can be seen in Fig~\ref{fig:haldane} (a), for weak disorder the Chern insulator is still characterized by a gap in the DOS where the Hall conductivity is quantized ($\sigma_{xy}=e^2/h$).  For increasing values of $T$, the longitudinal and transverse resistivities are in qualitative agreement with the experimental results of Ref. [\onlinecite{chern_science}],  with the suppression of both the peak in $\rho_{xy}$ and the dip in $\rho_{xx}$, supporting their findings.

A Chern insulator with a band gap $\Delta$ can be obtained by either having $\Delta_{AB}=0$ with $\Delta_T=\Delta$ or $\Delta=\Delta_T-\Delta_{AB}$. In both situations, the QAHE leads to $\sigma_{xy}=e^2/h$ that survives to intermediate disorder strength. Surprisingly, the two systems respond differently to strong disorder: as can be seen in Fig.~\ref{fig:haldane}, while disorder closes the gap and destroys the Chern insulator in the system with inversion symmetry (panel b), the QAHE with $\Delta_{AB}\neq0$  is insensitive to increasing Anderson disorder (panel c).  As illustrated in panel (c), large disorder can localize carriers and extend the topological phase to energies in the vicinity of the bulk gap,  similarly to what is observed in topological Anderson insulators~\cite{topological_anderson,topological_anderson-dis}.  For illustration purposes, the values of $\Delta_T$ and $\Delta_{AB}$ used in panel (c) are large in comparison with values in (b). However, the same effect can be seen if  $\Delta_T> \Delta_{AB}\neq 0$. 
To understand this behavior, we need to compare the gaps at the Dirac points in these two situations:  For $\Delta_{AB}=0$, the two valleys are degenerate and  the gaps in $K$ and $K^\prime$ are both $\Delta_T$.  On the other hand, for  $\Delta_{AB}\neq 0$, the interplay between $\Delta_T$  and  $\Delta_{AB}$ lifts the degeneracy between valleys so that one has $\Delta=\Delta_T-\Delta_{AB}$ and the other has  $\Delta=\Delta_T+\Delta_{AB}$ (see panel d). The gap difference has important consequences for the transport properties of the system. For $E_F$ in the range $\Delta_T+\Delta_{AB}>E>\Delta_T-\Delta_{AB}$, all the states belong to $K$ (the point group symmetry is $C_{3}$) and inter-valley scattering is forbidden as there are not available states connected to  $K^\prime$. This situation results in a smaller longitudinal resistivity. Also, it protects the topological gap and the QAHE as inter-valley scattering is detrimental to the state. Counter-intuitively, an asymmetry between sublattices A and B can help to stabilize the Chern insulator. In the limit of  $\Delta_T=\Delta_{AB}$ the gap closes in one of the valleys, producing a state that is protected from intervalley scattering and emulates a Weyl semimetal.

In summary, we have developed a numerical method to calculate the longitudinal and transverse conductivities of tight-binding hamiltonians in real space.  We illustrated the stability of the method by applying it to the QHE in disordered graphene, studying how the method's accuracy varies with the number of moments used in the expansion. To further illustrate the power of the method, we investigated the effect of disorder in the transport properties of a Chern insulator and found that due to the suppression of intervalley scattering, a Chern insulator with broken inversion symmetry is protected against scalar disorder. This finding can be useful in the search of Chern and topological insulating phases in novel materials.

The technique we have described is very general, and is suitable for the calculation of transport properties in finite temperature, disordered systems. One can simulate very large system sizes due to the method's high parallelizability that can be exploited in GPUs. Among other systems, we envisage that this method will be useful in the study of novel models with non-trivial topologies~\cite{dasasarmalattice}, spin transport in topological insulators, as well as materials without a topological phase, such as spin Hall conductivity in graphene. It can also be easily adapted to different geometries and multilayers of different materials.



\begin{acknowledgments}

We acknowledge A. R. Hernandez, A. Ferreira and E. Mucciolo for discussions.   T.G.R and J.H.G
acknowledge the Brazilian agencies CNPq, FAPERJ and INCT de Nanoestruturas de Carbono for financial
support. L.C. acknowledges the Flemish Science Foundation (FWO-Vlaanderen) for financial support.
\end{acknowledgments}


\newpage

\onecolumngrid

\textbf{ Supplementary Material for ''Real-space calculation of the conductivity tensor for disordered topological matter''}


In this supplementary material, we show in detail how the kernel polynomial method (KPM) can be used
to calculate $\sigma_{\alpha,\beta}(\mu)$  in an efficient and scalable way for all energies and
temperatures. For consistency,  let us first review some basic information about the KPM. This
method works by expanding a spectral operator of interest $\Lambda(\varepsilon,H)$  in terms of  the
Chebyshev polynomials of first kind $$T_m(x)\equiv \cos(m\text{arcos}(x)),\quad T_{m+1}(x)$$. For this purpose,  the
Hamiltonian needs to be rescaled so that its energy spectrum is contained in  the [-1,1] interval. 
This can be done by rescaling the Hamiltonian and  energies in the following way:
\begin{align}
\tilde{H}=\frac{2}{\Delta E}\left(H -\frac{E^{+}+E^{-}}{2}\right),\quad 
\tilde{\varepsilon}=\frac{2}{\Delta E}\left(\varepsilon -\frac{E^{+}+E^{-}}{2}\right),
\end{align}
where $E^{+}(E^-)$ is the higher (lower) bound of the spectra and $\Delta E = E^{+}-E^-$. To expand the scaled spectral quantity  $\tilde{\Lambda}(\tilde{\varepsilon},\tilde{H})$ we use the spectral representation 
\begin{align}
\tilde{\Lambda}(\tilde{\varepsilon},\tilde{H})=\sum_{k}\tilde{\Lambda}(\tilde{\varepsilon},\tilde{E}
_k)\ket{\tilde{E}_k}\bra{\tilde{E}_k}\label{SpectralMatrix},
\end{align}
where $\tilde{H}\ket{\tilde{E}_k}=\tilde{E}_k\ket{\tilde{E}_k}$  and expand each of the functions as follows:\cite{RMP_KPM_2006}
\begin{align}
\tilde{\Lambda}(\tilde{\varepsilon},\tilde{E}_k)=\frac{2}{\pi}\sum_{m=0}^{\infty}
\Gamma_m(\tilde{\varepsilon}) T_m(\tilde{E}_k),\quad
\Gamma_m(\tilde{\varepsilon})=\frac{1}{\delta_{m,0}+1}\int_{-1}^{1}\frac{\tilde{\Lambda}(\tilde{
\varepsilon},\tilde{E}_k)T_m(\tilde{E}_k)}{\sqrt{1-\tilde{E}_k^2}}d\tilde{E}_k,
\end{align}
where the Chebyshev polynomials can be efficiently evaluated by means of the recursion
relationship $T_m(x)=2xT_{m-1}(x)-T_{m-2}(x)$.  Finally, by inserting the above expression into
\eq{SpectralMatrix} we obtain the expansion for the spectral operator
\begin{align}
\tilde{\Lambda}(\tilde{\varepsilon},\tilde{H})=\frac{2}{\pi}\sum_{m=0}^{\infty}
\Gamma_m(\tilde{\varepsilon}) T_m(\tilde{H})\label{FinalSpectralMatrix},
\end{align}
where now all the information associated to the form of $\Lambda$ had been separated from the
information associated with the Hamiltonian. This type of expansion was used to approximate spectral
operators such as the Greens functions  \cite{PRL_GREEN_2006} and the evolution
operator\cite{PLA_EVOLUTION_2009,PRA_EVOLUTION_1998}. In many cases , only part of the information
contained in the Hamiltonian is needed. This is for example the case of the partition function
$\mathcal{Z}$ and  the density of states $\rho(\varepsilon)$\cite{JCP_DOS_1996,RMP_KPM_2006}, which
depends only on the trace of $H$. As an example, we can consider 
$\tilde{\rho}(\tilde{\varepsilon})$ where
$\tilde{\Lambda}(\tilde{\varepsilon},\tilde{H})=\delta(\tilde{H}-\tilde{\epsilon})$ and
\eq{FinalSpectralMatrix} becomes

\begin{align}
\tilde{\rho}(\tilde{\varepsilon})&=\frac{2}{\pi\sqrt{1-\tilde{\varepsilon}^2}}\sum_{m=0}^{\infty}
\Gamma_m(\tilde{\varepsilon}) \mu_m, \label{FinalSpectralScalar}
\end{align}
where  $\Gamma_m(\tilde{\varepsilon})=T_m(\tilde{\varepsilon})$ and
$\mu_m\equiv\text{Tr}[T_m(\tilde{H})]$, obtained in \onlinecite{JCP_DOS_1996} by expanding the
$\delta$-function directly.  The equation \eq{FinalSpectralScalar} is the general form of the trace
of an arbitrary spectral operator and the expansion is exact. However,  in practice it is always
necessary to truncate the series at some finite order $M$ and experience shows that this truncation
can produce poor precision and Gibbs oscillations, specially in points where th efunction is not
continuously differentiable.  One way to reduce finite order problems is to modify the moments 
$\mu_m\rightarrow g_m \mu_m$ with the use of a kernel, that  basically smooth the problematic
points. The Jackson Kernel,
\begin{align}
g_m^J=\frac{(M-m+1)\cos\frac{\pi m}{M+1}+\sin\frac{\pi m}{M+1}\text{cot}\frac{\pi}{M+1}}{M+1},
\end{align}
has been extensively  tested and studied  \cite{MAT_JACKSON_1930, LNP_JACKSON_2008,RMP_KPM_2006} and it is one of the best kernels for spectral quantities, as it has the advantage of being positive and normalized. Consequently, it preserves the positiveness and the value of the integration of the approximated function.  Calculating the trace of all $\text{Tr}[T_m(H)]$ can be a difficult task for large matrices. Fortunately, 
 there is a method, known as random phases vector approximation \cite{RMP_KPM_2006,PRL_DRABOLD_1993,IJMPC_SILVER_1994},  to calculate the traces that takes advantage of the size $N$ of the matrices . In this method we construct a set of vectors of $R\ll N$ complex vectors

\begin{align}
\ket{r}\equiv(\xi^r_1 ,\dots,\xi^r_N),\quad r=1,\dots,R,
\end{align}
 with dimension  equal to $N$ and whose elements $\xi^r_i$ are drawn from a probabilistic distribution with the following characteristics: 
\begin{align}
\medio{\medio{\xi^r_i}}=0,\quad
\medio{\medio{{\xi^r_i}^*\xi^{r'}_j}}=\delta_{r,r'}\delta_{i,j}\label{conditions},
\end{align}
where $\medio{\medio{\dots}}$ is the statistical average.  The trace can be calculated as the average expected value of this  random vector,
\begin{align}
\text{Tr}\left[T_m(\tilde{H})\right]\approx \frac{1}{R}\sum_{r=1}^R \bra{r}T_m(\tilde{H})\ket{r}.
\end{align}

The error of this approximation  is $\mathcal{O}(1/\sqrt{RN})$, therefore it  is reduced when increasing $N$ or $R$ and for very large system only a few random vectors are necessary. In principle, it is possible to use any distribution function that satisfies the condutions of Eq. \eq{conditions}. However, it was prove that choosing $\xi^r_i=\exp{i\phi}/N$, with $\phi$ as a random variable uniformly distributed in the interval $(0,2\pi)$ reduces the statistical error, therefore the convergence is improved \cite{PRE_IITAKA_2004,IJMPC_SILVER_1994}.

Now that we presented the basics of the KPM, we can proceed to expand the conductivity tensor $\sigma_{\alpha,\beta}(\mu)$. Our starting point is the the formula given by  Bastin {\it et al.} \cite{JPC_BASTIN_1971} with the assumption that the Hamiltonian has finite lower and upper bounds, which is always true for a tight-binding Hamiltonian:
\begin{align}
&\sigma_{\alpha\beta}(\mu)= \frac{ie^2\hbar}{\Omega}
\int_{E^{-}}^{E^{+}}d\varepsilon f(\varepsilon) \text{Tr}\left\langle v_\alpha\delta(\varepsilon-H)v_\beta \d{G^{+}(\varepsilon,H)}{\varepsilon}-v_\alpha  \d{G^{-}(\varepsilon,H)}{\varepsilon}v_\beta \delta(\varepsilon-H)\right \rangle. \label{Sigma}
\end{align} 
Here, $\Omega$ is the volume, $G^{\pm}(\epsilon,H)$ are the advanced and retarded Green functions, $v_\alpha$  is the velocity operator in the $\alpha$ direction and $f(\varepsilon)$ is the Fermi-Dirac distribution for a given chemical potential $\mu$ and temperature $T$.  We begin by rescaling of the Hamiltonian $H\rightarrow\tilde{H}$ and the energies $\varepsilon\rightarrow\tilde{\varepsilon}$  and replace them in the Bastin's Formula
\begin{align}
\sigma_{\alpha\beta}(\mu)&= \left(\frac{2}{E^{+}-E^{-}}\right)^2  \frac{ie^2\hbar}{\Omega}
\int_{-1}^{1}d\tilde{\varepsilon} f(\tilde{\varepsilon},\tilde{\mu}) \text{Tr}\left\langle
v_\alpha\delta(\tilde{\varepsilon}-\tilde{H})v_\beta
\d{G^{+}(\tilde{\varepsilon},\tilde{H})}{\tilde{\varepsilon}}-v_\alpha 
\d{G^{-}(\tilde{\varepsilon},\tilde{H})}{\tilde{\varepsilon}}v_\beta
\delta(\tilde{\varepsilon}-\tilde{H})\right \rangle \nonumber\\
\sigma_{\alpha\beta}(\mu)&=\left(\frac{2}{E^{+}-E^{-}}\right)^2  \tilde{
\sigma}_{\alpha\beta}(\tilde{\mu}),
\end{align}
so we can work  with the rescaled conductivity $\tilde{ \sigma}_{\alpha,\beta}(\tilde{\mu})$.  Using the expression \eq{FinalSpectralMatrix}, we can expand both the $\delta$ and the Green function in terms of the Chebyshev polynomials
\begin{align}
\delta(\tilde{\varepsilon}-\tilde{H})=\frac{2}{\pi\sqrt{1-\tilde{\varepsilon}^2}}\sum_{m=0}^{\infty} T_m(\tilde{\varepsilon})T_m(\tilde{H}),\quad 
G^{\pm}(\tilde{\varepsilon},\tilde{H})=\mp \frac{2i }{\sqrt{1-\tilde{\varepsilon}^2}}\sum_{m=0}^{M}g_m \exp{\pm i m \text{arccos}(\tilde{\varepsilon})}T_m(\tilde{H}).
\end{align}
By replacing the functions above in \eq{Sigma} we have 
\begin{align}
\tilde{\sigma}_{\alpha\beta}(\tilde{\mu})&=  \frac{4 e^2\hbar}{\pi\Omega}  \int_{-1}^{1}
\frac{d\tilde{\varepsilon}~
f(\tilde{\varepsilon})}{(1-\tilde{\varepsilon}^2)^2}\sum_{m,n}\mu_{nm}^{\alpha\beta}
\Gamma_ {nm}(\tilde{\varepsilon})\label{condfinal},
\end{align}
where 
\begin{equation}
\mu_{mn}^{\alpha\beta}\equiv \frac{g_m g_n}{(1+\mu_{n0})(1+\mu_{m0})}\text{Tr}\left[v_\alpha
T_m(\tilde{H})v_\beta T_n(\tilde{H})\right]\label{Moments},
\end{equation}
 does not depend of $\tilde{\varepsilon}$ and carries all the information about the system, being
also responsible for most of the computational cost.  On the other hand, 
\begin{align}
\Gamma_{mn}(\tilde{\varepsilon})\equiv&\left(\varepsilon -i n \sqrt{1-\varepsilon ^2}\right)
T_m(\varepsilon ) e^{i n \text{arccos}(\varepsilon )}+
\left(\varepsilon +i m \sqrt{1-\varepsilon ^2}\right)  T_n(\varepsilon ) e^{-i m
\text{arccos}(\varepsilon )},\end{align}
is independent of the Hamiltonian  and can be thought as the expansion basis. \eq{condfinal} can be seen as a generalization of  \eq{FinalSpectralScalar} where more than two spectral functions are present in the expansion.  By using the properties 
\begin{align}
(\mu_{mn}^{\alpha\beta})^*=\mu_{nm}^{\alpha\beta}=\mu_{mn}^{\beta\alpha}\quad\text{and}\quad
\Gamma_{mn}^*=\Gamma_{nm}\label{condproperties},
\end{align}
it is possible to write the conductivity as a complete real quantity
\begin{align}
\tilde{\sigma}_{\alpha\beta}(\mu)&= \frac{4 e^2\hbar}{\pi\Omega} \int_{-1}^{1}
\frac{d\tilde{\varepsilon}~ f(\tilde{\varepsilon})}{(1-\tilde{\varepsilon}^2)^2}\sum_{m,n\leq
m}\text{Re}\left[\mu_{nm}^{\alpha,\beta}\Gamma_{nm}(\tilde{\varepsilon})\right].\label{
condfinalSymm}
\end{align}

Finally, as an example of the consistency of our approach, we integrate this equation analytically
to obtain the Kubo-Greenwood formula for the longitudinal conductivity.  For this purpose we can use
\eq{condproperties} to show that $\mu_{mn}^{\alpha\alpha}$ is  real  and therefore
$\text{Re}\left[\mu_{nm}^{\alpha\alpha}\Gamma_{n,m}(\tilde{\varepsilon})\right]=\mu_{nm}^{\alpha
\alpha}\text{Re}\left[\Gamma_{nm}(\tilde{\varepsilon})\right]$. Taking this into account, we can
write
\begin{align}
\tilde{\sigma}_{\alpha\alpha}(\tilde{\mu})&= \frac{4 e^2\hbar}{\pi\Omega} \sum_{m, n\leq
m}\mu_{nm}^{\alpha\alpha}  \int_{-\alpha}^{\alpha} \frac{d\tilde{\varepsilon}~
f(\tilde{\varepsilon})}{(1-\tilde{\varepsilon}^2)}\left(
\d{}{\tilde{\varepsilon}}\left[\text{T}_n(\tilde{\varepsilon})
\text{T}_m(\tilde{\varepsilon})\right]+2 \tilde{\varepsilon}
\text{T}_n(\tilde{\varepsilon})\text{T}_m(\tilde{\varepsilon})\right),
\end{align}
that  can be integrated analytically for $T=0$:
\begin{align}
\sigma_{\alpha\alpha}(\tilde{\varepsilon}_F)&= \frac{4 e^2\hbar}{\pi\Omega} \sum_{m, n\leq
m}\mu_{nm}^{\alpha\alpha}  T_n(\tilde{\varepsilon}_F)T_m(\tilde{\varepsilon}_F),
\end{align}
which is the Kubo-Greenwood formula  expressed in terms of the Chebyshev Polynomials as shown in  \cite{RMP_KPM_2006}.

\subsection{Convergence of the method} 

\begin{figure}[h!]
     \centering\includegraphics[width=0.7\linewidth,clip]{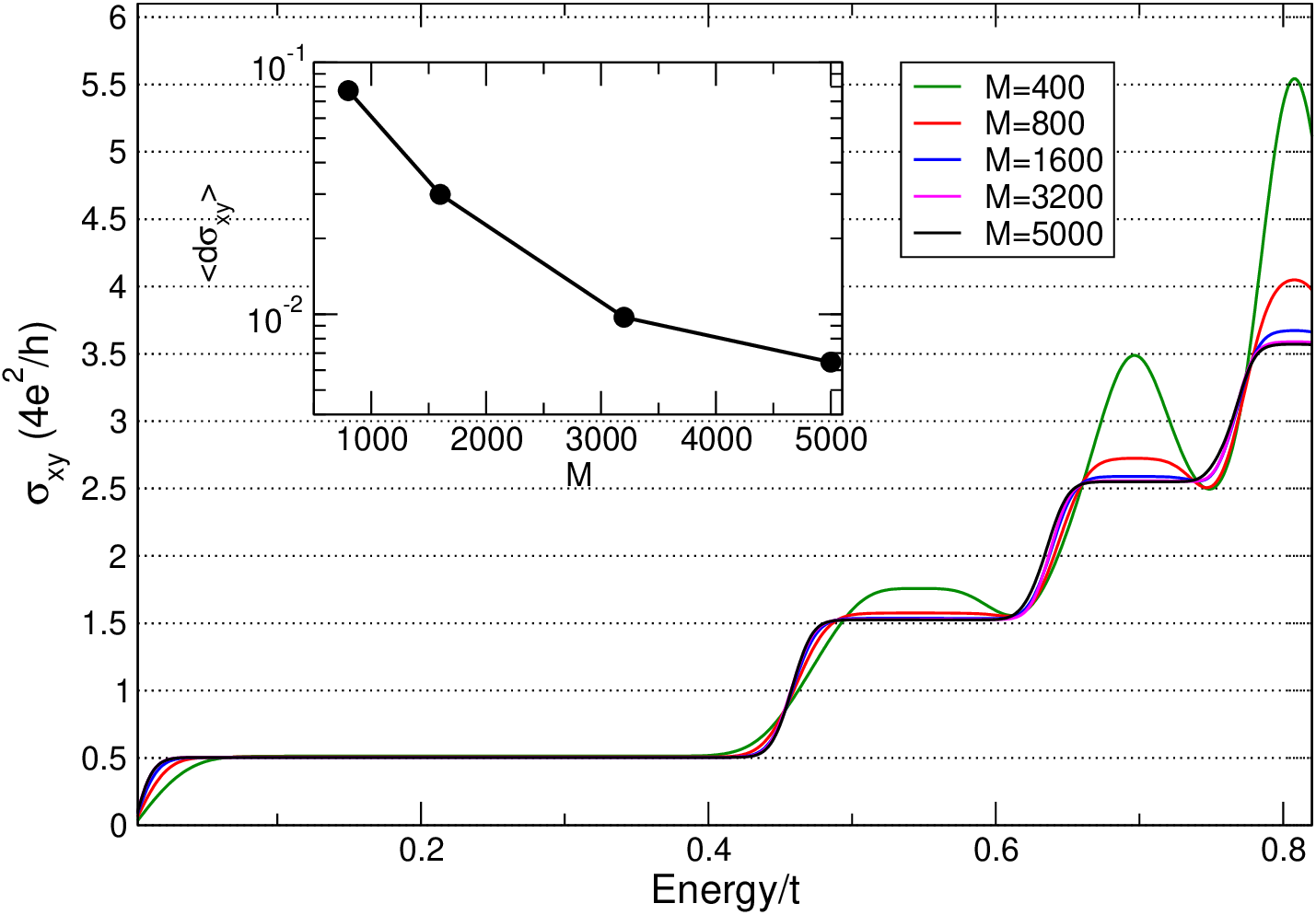}
     
      \caption{ $\sigma_{xy}$ as a function of E for increasing values of $M$. $\phi/\phi_0\approx 1\times 10^{-2}$, $SR>150$, $K_BT=0.005t$ and $N=2\times 400\times 400$ sites. Inset: ${ \langle d\sigma_{xy} (E) \rangle}$ as a function of $M$ for the data shown in the main panel. }
         \label{fig:hall}
\end{figure}

In the figure 1, we illustrate how we track the convergence in function of $M$: in this particular case,  we choose the window $(0,0.8t]$ in energy and calculate $\sigma_{xy}(E)$ for two different values of $M$ ( $M^\prime < M$). We then obtain $${ \langle d\sigma_{xy} (E) \rangle}= \left \langle  \left | \frac{\sigma^M_{xy}(E)-\sigma^{M^\prime}_{xy}(E)}{\sigma_{xy}^M(E)}  \right | \right \rangle_E,$$ where we calculate the discrepancy  between the conductivity for different values of $M$ at each energy point and average over the energy.  In the inset, we can see that for $M=5000$,  ${ \langle d\sigma_{xy} (E) \rangle}<10^{-2}$ which means that further increases in the number of moments of the expansion will provide very small changes in the conductivity.  The conductivity converges slowly for large values of $M$ ( ${ \langle d\sigma_{xy} (E) \rangle}\propto M^{-1.4}$) and for increasing values of $M$, the changes will mostly occur at high energies. If we choose a different window of energy, as for example  $(0,0.4t]$,  ${ \langle d \sigma_{xy} (E) \rangle}<10^{-3}$ even for $M=3200$.  We then truncate our expansion for ${ \langle d\sigma_{xy} (E) \rangle}\sim10^{-2}$.

A similar analysis can be performed with the number of random vectors $R$, as shown in figure 2. For $R$ , we consider ${ \langle d\sigma_{xy} (E) \rangle}\sim10^{-3}$ . {\it{Our general procedure is to converge $SR$ for each value of $M$ for a given energy range and then look at the convergence as a function of $M$}}. It is important to mention that for increasing values of $M$, convergence requires increasing values of $SR$.  However, temperature reduces small fluctuations in the conductivity that arise from the use of few random vectors, eliminating the need of large $SR$ for finite $T$.

\begin{figure}[h!]
     \centering\includegraphics[width=0.4\linewidth,clip]{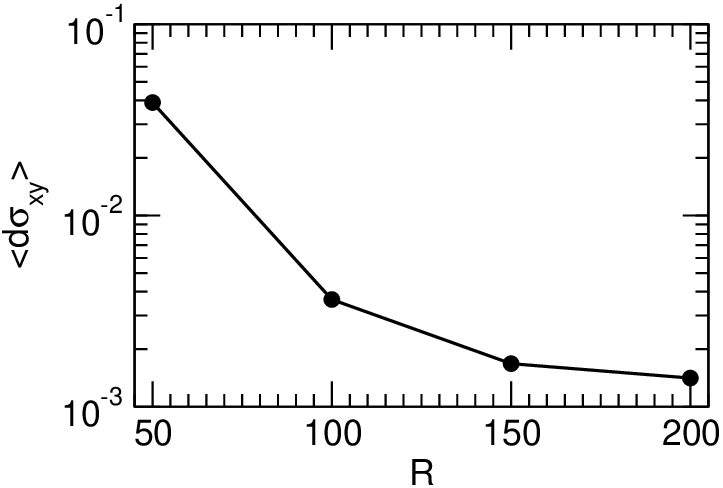}
     
      \caption{ (a)   ${ \langle d\sigma_{xy} (E) \rangle}$ as a function of $R$ for $\phi/\phi_0\approx 1\times 10^{-2}$, $M=3200$, $k_BT=0$ and $N=2\times 400\times 400$ sites. }

         \label{fig:hall}
\end{figure}

The Hall conductivity presents deviations from the quantized value for high energies. However, if we look at the values of the quantized changes in $\sigma_{xy}$  in the figure below ($M=5000$), for positive $n$ we have 1.015, 1.025, 1.03, 1.03, 1.06  in units of $ 4e/h^2$. The contribution of each Landau level to the Hall conductivity has an error or the order of 1-5 $\%$. However, the error in the conductivity accumulates and deviates from the expected values at high LL levels. 
The small increase of the error with energy is related with the convergence with $M$ discussed above.

\subsection{GPU Computing}

As discussed  in the previous section, the main computational cost of our approach is the
calculation of elements such as  $T_m(\tilde{H})\ket{r}$ with
$T_m(\tilde{H})=2\tilde{H}T_{m-1}(\tilde{H})-T_{m-2}(\tilde{H})$. For a conductivity calculation
with $M=10^3$ it is necessary to calculate around $10^6$  matrix-vector products, with matrices of
dimension $N\times N$ and $N\approx10^5-10^7$.  To reduce the computational time, we take advantage
of the fact that the real space tight-binding Hamiltonians are represented by sparse matrices, so we
write the matrix in a sparse format which greatly reduces the computational cost from
$\mathcal{O}(N^2)\rightarrow \mathcal{O}(N)$.  We take advantage of the parallel nature of  Graphics
Processing Units (GPU) to compute these products in a very efficient way. We perform our simulations
in CUDA,   NVIDIA  proprietary parallel computing platform. We also use NVIDIA CUDA Sparse Matrix
library (cuSPARSE) that provides  basic linear algebra subroutines used for sparse matrices and
CUSP, another Sparse Matrix library for CUDA . The performance of these packages had been tested by
NVIDIA and achieves an speedup of roughly 10 times of a common CPU with MKL optimized
libraries\cite{CUDA}.

\end{document}